\documentclass[runningheads]{llncs}
\usepackage[T1]{fontenc}
\usepackage{graphicx}
\usepackage{booktabs}
\usepackage{array}
\usepackage{xcolor}
\usepackage[table]{xcolor}
\usepackage{float}
\usepackage[multiple]{footmisc}
\usepackage{fancyhdr}

\renewcommand{\arraystretch}{1}
\usepackage[margin=1.2in]{geometry}

\newcommand{\OG}[1]{\textsc{OpenGadget3}}

\begin{document}
\title{Is RISC-V Ready for Massively Parallel Astrophysical  Codes?}

\author{Jenny Lynn Almerol \inst{1,2}\and
Nitin Shukla \inst{2}\and
Federico Ficarelli \inst{2}\and
Geray S. Karademir \inst{3} \and
Andrea Bartolini \inst{4}\and
Emanuele Venieri \inst{4}\and
Giacomo Madella \inst{4}\and 
Elisabetta Boella \inst{5}
}
\authorrunning{J. L. Almerol et al.}
%
\institute{
Scuola Internazionale Superiore di Studi Avanzati, Trieste, Italy \and
CINECA, Casalecchio di Reno, Italy \and
Universitäts-Sternwarte, Fakultät für Physik, Ludwig-Maximilians-Universität München, Munich, Germany \and
DEI, Università degli studi di Bologna, Bologna, Italy \and
E4 Computer Engineering SpA, Scandiano, Italy
}
\maketitle              
%

\thispagestyle{fancy}
\fancyfoot[C]{%
\parbox{\textwidth}{\footnotesize

{\centering
\textit{Pre-print version of the manuscript submitted to the RISC-V for HPC Workshop, ISC 2026.}
\par}

\vspace{0.2em}

\copyright\ 2026 IEEE. Personal use of this material is permitted. Permission from IEEE must be obtained for all other uses, in any current or future media, including reprinting/republishing this material for advertising or promotional purposes, creating new collective works, for resale or redistribution to servers or lists, or reuse of any copyrighted component of this work in other works.
}}

\begin{abstract}
We present a performance and portability evaluation of three well-established astrophysical production codes, namely iPIC3D, PLUTO, and \OG, on a Sophgo SG2044 RISC-V processor (part of the Monte Cimone cluster), with comparisons to AMD EPYC 9554 (x86) and NVIDIA GH200 Grace (ARM) systems. These applications represent memory-bound, compute-bound, and hybrid workloads, respectively. Numerical correctness is verified across all platforms, confirming portability. RISC-V shows consistently lower performance, with slowdowns of about $3-6\times$ relative to x86 and $5-9\times$ relative to ARM. The gap is mainly due to limited memory bandwidth, shared cache constraints, narrower 128-bit vector units, and lower clock frequency, but also less-mature auto-vectorization capability of the GNU compiler suite. Memory-bound kernels are the most affected, where early bandwidth saturation and L2 cache contention reduce scalability at higher thread counts.
Hybrid MPI+OpenMP configurations reveal a trade-off between memory contention and communication overhead, with intermediate configurations achieving the best performance. These results suggest that RISC-V is capable of supporting scientific workloads; however, additional improvements in both hardware and compiler technology, particularly in auto-vectorization, are required to achieve competitive performance.

\end{abstract}
\section{Introduction}
High Performance Computing (HPC) is entering an era of architectural diversification driven by the need for specialized computational capabilities, with heterogeneous systems integrating general-purpose and domain-specific processors (x86, ARM, RISC-V)~\cite{Hennessy2019}, massively parallel accelerators (GPUs)~\cite{simsek2024increasing}, reconfigurable logic (FPGAs)~\cite{silvano2025survey}, application-specific integrated circuits~\cite{Almerol2025_Wormhole,almerol2026riscv_nbody}, and emerging quantum computing technologies~\cite{elsharkawy2025integration,Viviani2025,Rocco_quantum}. Today supercomputers mostly rely on GPUs and proprietary CPU architectures\footnote{TOP500: \url{ https://www.top500.org/lists/top500/2025/11/}}, but there is a growing interest in open alternatives like RISC-V. These open architectures offer flexibility, reduce dependence on single vendors, and could reshape how we build future HPC systems. For the HPC community, the critical question is no longer whether RISC-V can execute scientific workloads, but whether it can support production-grade, massively parallel applications at scale~\cite{brown2025riscvreadyhighperformance}. In this regard, the Monte Cimone project~\cite{Venieri2026} helps the scientific community by deploying and operating an HPC cluster equipped with state-of-the-art RISC-V compute nodes.

Astrophysical simulation codes represent some of the most demanding workloads in scientific computing\footnote{\url{https://www.ornl.gov/news/record-breaking-run-frontier-sets-new-bar-simulating-universe-exascale-era}}. They combine multi-physics modeling, extreme dynamic range, irregular communication patterns, and long execution times, often consuming millions of core-hours per campaign. Ensuring that such applications remain portable and efficient across evolving architectures is essential for sustaining scientific productivity and protecting decades of software investment. Recent efforts to port astrophysical mini-applications, including 3D multi-physics and N-body solvers, to RISC-V have already demonstrated promising scalability~\cite{Diehl2023_OctoTiger} and, in some cases, performance exceeding optimized CPU implementations~\cite{Almerol2025_Wormhole}. While these results indicate a growing level of architectural maturity, a comprehensive assessment of flagship, production-grade applications remains necessary.

This work presents a systematic validation and performance evaluation of three flagship astrophysical applications, namely iPIC3D~\cite{Markidis2010_iPIC3D}, PLUTO~\cite{Berta2024} and \OG~\cite{Groth2023}, on the highest-performing commercially available RISC-V processor at the time of writing, deployed within the Monte Cimone cluster~\cite{Venieri2026}. These applications were previously optimized for x86-based architectures within the EuroHPC Joint Undertaking Centre of Excellence (CoE) SPACE\footnote{SPACE CoE: \url{https://www.space-coe.eu/}}, where strong scalability on systems using millions of CPU cores was demonstrated, with parallel efficiencies exceeding 80\%, indicating mature performance characteristics~\cite{Shukla2025_SPACE_CoE,Shukla2025_EuroHPCDay,Shukla2026_exascale}.

The objective of this work is therefore to evaluate the performance and portability of well optimized x86 codes to RISC-V platforms. Our study evaluates real production workloads spanning three complementary computational domains represented by these applications. In addition to quantitative performance measurements, the architectural readiness of RISC-V for production scientific computing is examined, and the remaining gaps in both hardware capabilities and the maturity of the software toolchain are identified. In summary, the main contributions of this paper are:
\begin{itemize}
\item Evaluation of the portability and performance of three astrophysical codes (iPIC3D, PLUTO, and \OG~) on a RISC-V platform, with comparisons with x86 and ARM.
\item Analysis of compiler-driven vectorization and identification of key performance bottlenecks, including memory bandwidth and cache behavior.
\item Study of hybrid MPI+OpenMP configurations on RISC-V, highlighting the impact of memory hierarchy on performance.
\end{itemize}


The paper is organized as follows: 
Section~\ref{testbed} describes the three testbed applications and their computational characteristics. Section~\ref{exp_setup} details the experimental setup, including the hardware platforms and the software stack. Section~\ref{results_and_discussion} presents results and discussion for each application. Section~\ref{summary} summarizes findings and outlines future directions.

\section{Testbed Applications}
\label{testbed}
Three different astrophysical production codes, part of the SPACE CoE, were selected for this study based on their maturity level, their widespread use within their respective domains, and the diversity of algorithms and computational workloads they represent.


iPIC3D\footnote{iPIC3D: \url{https://github.com/KTH-HPC/iPIC3D.git}} is a semi-implicit Particle-in-Cell (PIC) code developed to study collisionless plasma dynamics at kinetic scales~\cite{Markidis2010_iPIC3D,Williams2023_iPIC3D_SC23}. In this approach, macro-particles representing ensembles of plasma particles are advanced in a Lagrangian framework, while plasma moments such as density, current, and pressure, together with the self-consistent electric and magnetic fields, are evaluated on a Eulerian grid. The code is written in C/C++ and parallelized via MPI. Accelerated versions leveraging OpenACC and CUDA have been recently developed~\cite{BoellaGPU,MarkidisGPU}. A version of the code integrated with the Legio fault resilience framework also exists~\cite{Rocco2025}.
The computational structure of iPIC3D is dominated by three main kernels: the Particle Mover, the Moment Gatherer (where particles are interpolated to the grid), and the Field Solver. 



PLUTO\footnote{PLUTO: \url{https://gitlab.com/PLUTO-code/gPLUTO}} is a widely used code for computational plasma astrophysics, designed to solve the equations of magnetohydrodynamics (MHD) in multiple spatial dimensions using a finite-volume formulation~\cite{Mignone2007_PLUTO,Mignone2010_GLM_MHD,Rossazza2026_gPLUTO,Suriano2026_LP}. It is implemented primarily in C (with C++ required for the Adaptive Mesh Refinement interface) and solves MHD equations on structured grids in one, two, or three dimensions using multiple coordinate systems. The code is based on Godunov-type methods and supports high-resolution simulations of nonlinear systems of conservation laws, making it suitable for modeling complex phenomena such as accretion disks, jets, and turbulent plasma flows. It is highly scalable, supporting execution from single workstations to large distributed-memory systems via MPI, and includes a comprehensive set of fluid-dynamical test problems for benchmarking, along with a Python-based analysis and visualization tool (pyPLUTO).




\OG~(Dolag et al. in prep) is an astrophysical simulation code written in C and C++ aimed at large full-physics cosmological simulations. The code calculates gravitational forces using an N-body approach, combining an oct-tree Barnes-Hut algorithm~\cite{BarnesHut1986} for small scales and a particle mesh method approach utilizing the Fastest Fourier Transform in the West (FFTW3) for large ranges. The hydrodynamic density and forces are computed using smoothed particle hydrodynamics (SPH). As an alternative to SPH, the code also allows the use of a meshless finite mass method (MFM)~\cite{Groth2023} to solve the hydrodynamical equations. In addition, the code includes baryonic physics processes such as radiative cooling, star formation, energy feedback, radiative transfer, magnetic fields, and black hole evolution. \OG~ employs a hybrid MPI + OpenMP parallelization strategy, using a Hilbert space-filling curve to optimize domain decomposition and data locality.

The three astrophysical codes exhibit complementary computational characteristics
and provide a basis to evaluate RISC-V CPUs. iPIC3D is largely memory-bound, dominated by particle-to-grid and grid-to-particle operations, with its implicit solver reducing communication frequency. PLUTO is compute-bound with structured memory access that benefits from vectorization, while MPI halo exchanges affect parallel efficiency. \OG~ combines memory- and latency-sensitive tree computations with compute-heavy particle-mesh FFTs and SPH hydrodynamics, employing hybrid MPI + OpenMP and Hilbert curve decomposition to optimize load balance and communication. Together, these codes enable assessment of numerical correctness, parallel scalability, compiler optimization, and memory bandwidth utilization on RISC-V platforms.
We take advantage of the distinct computational characteristics of iPIC3D, PLUTO, and \OG~ to establish a comprehensive evaluation framework for RISC-V within production-grade HPC environments. 


\section{Experimental Setup}
\label{exp_setup}
Benchmarks were conducted on three compute platforms spanning three distinct instruction set architectures (ISAs): x86, AArch64 (ARM), and RISC-V. Cross-architecture performance comparisons use the Sophgo SG2044 as the primary RISC-V platform.

\textit{Hardware Platforms.}
The benchmark suite was executed across three distinct environments to provide a cross-architecture comparison. The x86 node, hosted at E4 Computer Engineering, utilizes a dual-socket configuration featuring the AMD Genoa microarchitecture. These sockets are linked via AMD Infinity Fabric to manage the 128 physical cores.

The ARM node is an NVIDIA GH200 Grace Hopper Superchip, also located at E4 Computer Engineering. While this platform features a tight integration between the CPU and GPU via NVLink-C2C, this study utilized only the CPU cores to provide a direct architectural comparison with x86 and RISC-V platforms. The 72 Neoverse-V2 cores operate as a single unified CPU NUMA domain. Although the operating system reports additional NUMA nodes, these correspond exclusively to the GPU HBM3 memory regions and were not engaged during the benchmarks.

The RISC-V node, hosted at the University of Bologna as part of the Monte Cimone cluster~\cite{bartolini2022monte,montecimone2}, is powered by the Sophgo SG2044 processor. This architecture organizes its 64 T-Head C920v2 cores into clusters of four. Each cluster shares a 2 MiB L2 cache, while all cores share a 64 MiB L3 cache and reside in a single NUMA domain. Detailed specifications for all three systems are provided in Table~\ref{tab:hardware}.

\begin{table}[tp!]
\caption{Hardware specifications of the three compute platforms used in this study.}
\vspace{-0.2cm}
\label{tab:hardware}
\centering
\small
\setlength{\tabcolsep}{6pt}
\renewcommand{\arraystretch}{1.0}

\begin{tabular}{p{3.5cm} p{3cm} p{4cm} p{2.9cm}}
\toprule
\rowcolor{gray!15}
\textbf{Specification} & \textbf{x86} & \textbf{ARM} & \textbf{RISC-V} \\
\midrule

\textbf{System / Chip} 
& AMD EPYC 9554 
& NVIDIA GH200 (Grace) 
& Sophgo SG2044 \\

\textbf{Microarchitecture} 
& AMD Genoa 
& ARM Neoverse-V2 
& T-Head C920v2 \\

\textbf{ISA} 
& x86 
& AArch64 
& RV64GC \\

\textbf{Sockets} 
& 2 
& 1 
& 1 \\

\textbf{Total Cores} 
& 128 
& 72 
& 64 \\

\textbf{Threads / Core} 
& 1 
& 1 
& 1 \\

\textbf{Peak Clock (GHz)} 
& 3.76 
& 3.47 
& 2.50 \\


L1-D / L1-I Cache 
& 32~KiB / 32~KiB 
& 64~KiB / 64~KiB 
& 64~KiB / 64~KiB \\

L2 Cache 
& 1~MiB/core 
& 1~MiB/core 
& 2~MiB/4-core cluster \\

L3 Cache 
& 32~MiB 
& 114~MiB 
& 64~MiB (shared) \\


NUMA Domains (CPU) 
& 2 (1 per socket) 
& 1 
& 1 \\

SIMD / Vector ISA 
& AVX-512, AVX2 
& SVE2 (128-bit) 
& RVV~1.0 (128-bit) \\

\bottomrule
\end{tabular}
\end{table}

\textit{Software Stack.}
All three codes use MPI for distributed-memory communication. \OG~ employs also OpenMP for
shared-memory thread-level parallelism on top of MPI. CMake serves as the build system for
all applications on every platform.

On the RISC-V node the entire software stack was built from source using
GCC~14.2.0, which provides full support for the finalized RVV~1.0 standard,
paired with OpenMPI~4.1.4. On the x86 and ARM nodes the environment consists
of GCC~12.4.0 and OpenMPI~5.0.6. All platforms share the same scientific
library versions: FFTW~3.3.10 (compiled with OpenMP support), HDF5~1.14.2
for parallel I/O, and GSL~2.7.1 for mathematical routines.

Application codes were compiled everywhere with \texttt{-O3 -ffast-math
-funroll-loops -ftree\-vectorize} and architecture-specific targets:
\texttt{-march=znver4} with full AVX-512 extensions on Zen~4;
\texttt{-march=armv9-a -msve-vector-bits=128}, enabling SVE at a fixed
128-bit width on Neoverse-V2; and \texttt{-march=rv64gcv
-mrvv-max-lmul=m8}, enabling RVV with LMUL\,=\,8 on the C920v2. The \texttt{-mrvv-\allowbreak max-lmul=m8} setting allows the
compiler to group up to eight 128-bit vector registers, so the
auto-vectorizer can group multiple vector registers and generate more compact instruction sequences.

\section{Results and Discussion}
\label{results_and_discussion}
We evaluated the behavior of the selected codes by running representative use cases on the target hardware platforms, with build and configuration settings described in Sect.~\ref{exp_setup}. To ensure statistical robustness, each simulation was repeated at least 10 times. In addition to performance measurements, we also analyzed compiler vectorization reports (generated via \texttt{-fopt-info-vec}) to better understand code generation in the different architectures. On the RISC-V machine, these reports confirmed that the regular loop structures in PLUTO and \OG~ were successfully auto-vectorized by GCC 14.2.0, generating RVV 1.0 instructions, whereas iPIC3D was not automatically vectorized under the same compilation settings.

\subsection{iPIC3D}
\label{sec:ipic_results}
\begin{figure}[tp]
\centering
\includegraphics[width=0.49\textwidth]{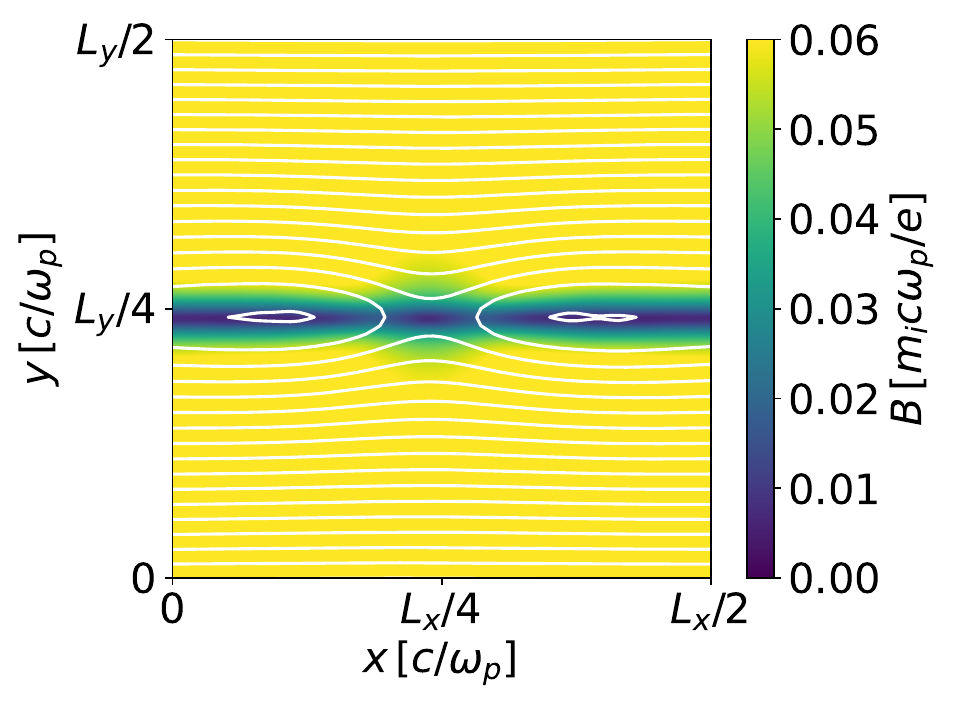}
\includegraphics[width=0.49\textwidth]{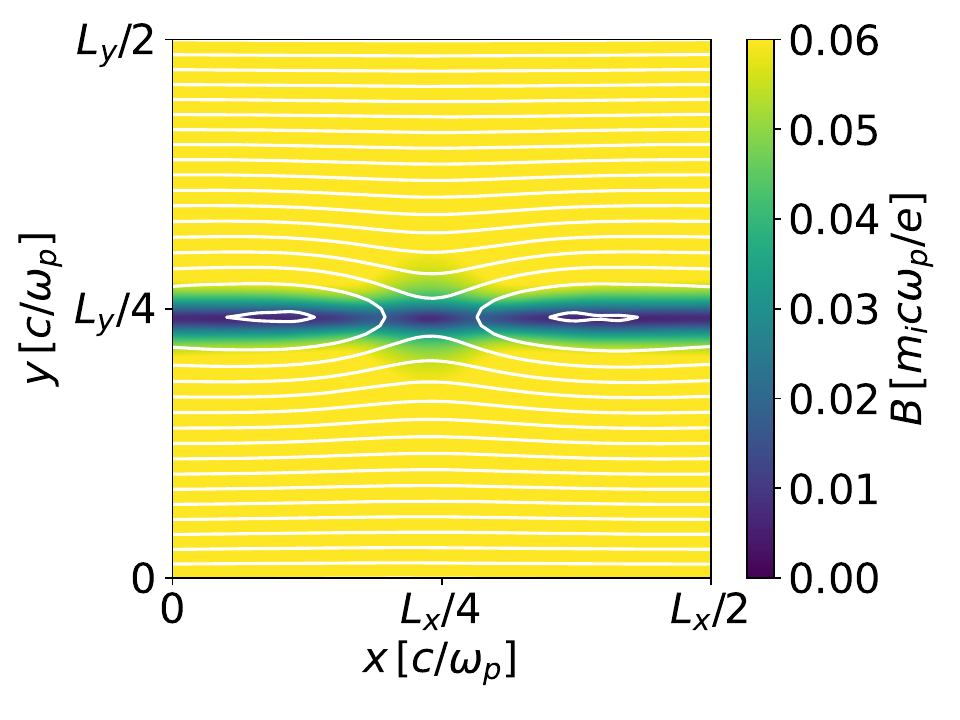}
\vspace{-0.7cm}
\caption{Magnitude of the magnetic field on the simulation plane (shown in color) with field lines overlaid in white, during an iPIC3D simulation of the GEM reconnection challenge (values are normalized to reference quantities). \textbf{Left:} run performed on SG2044. \textbf{Right:} run performed on GH200.}
\label{fig:ipic_validation}
\end{figure}

Before assessing the performance of iPIC3D, we first verified the correctness of its results on the SG2044 platform. To this end, we simulated the use case provided by the Geospace Environmental Modeling (GEM) reconnection challenge~\cite{GEM} on both RISC-V and ARM-based nodes. The results, shown in Fig.~\ref{fig:ipic_validation}, present the magnitude of the magnetic field in the simulation plane along with the corresponding field lines at a stage where magnetic reconnection is actively occurring. The close agreement between the two sets of results confirms the validity of the simulation performed on the RISC-V architecture.

We proceeded to evaluate iPIC3D employing the same setup on a $256\times256\times1$ spatial grid with 9216 particles per cell, resulting in a problem size large enough to fully utilize the available memory of each node. The simulation was evolved for 5 cycles with a pure-MPI decomposition using all 64 cores of the RISC-V node, and equivalently configured on the ARM and x86 nodes for cross-architecture comparison.

\begin{figure}[t]
\centering
\includegraphics[width=0.49\textwidth]{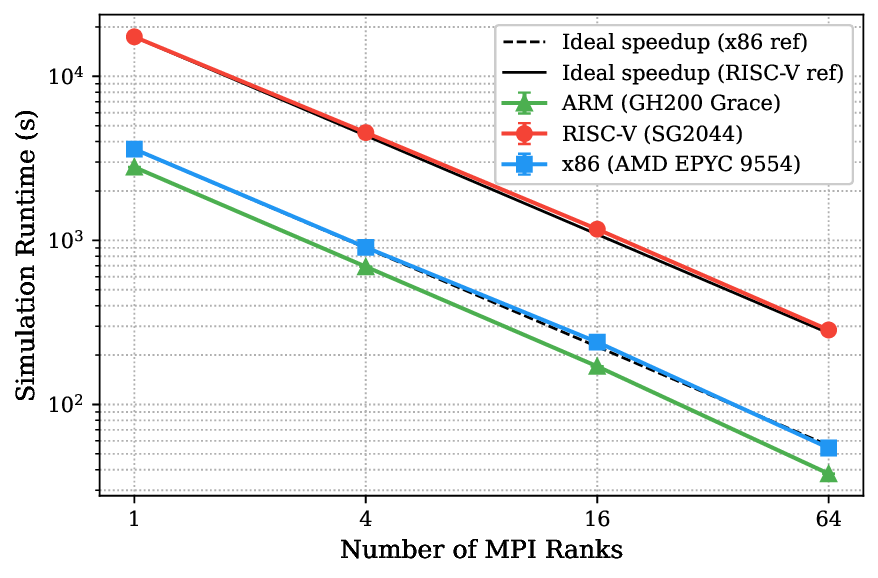}
\hfill
\includegraphics[width=0.49\textwidth]{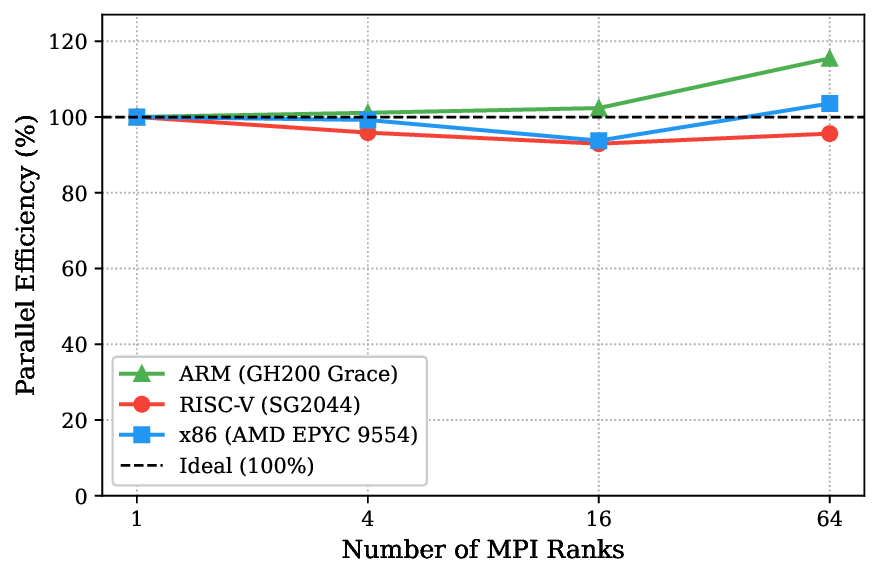}
\vspace{-0.5cm}
\caption{iPIC3D strong-scaling results for a $256\times256\times1$ grid and 9216 particles per cell on x86, ARM, and RISC-V platforms. \textbf{Left:} wall clock time versus MPI rank count (lower is better). \textbf{Right:} parallel efficiency normalized to the single-rank execution time on each platform.}
\label{fig:ipic_ss_pe}
\end{figure}

The baseline iPIC3D application uses an Array-of-Structures (AoS) memory layout for particle data, where per-particle quantities are interleaved in memory. This layout results in non-unit-stride access and prevents the compiler auto-vectorizer from generating RVV instructions for the performance-critical inner loop. To investigate the impact of memory layout on RISC-V performance, additional experiments were conducted comparing three configurations: unsorted AoS (the baseline), unsorted Structure-of-Array (SoA), and sorted AoS. Sorting particle arrays renders the innermost scatter access pattern sequential, enabling compiler auto-vectorization of the moment gathering kernel. As alternative configurations require increased memory usage, the problem size was reduced to 6400 particles per cell for these experiments to avoid out-of-memory (OOM) errors.

\begin{figure}[t]
\centering
\includegraphics[width=0.75\textwidth]{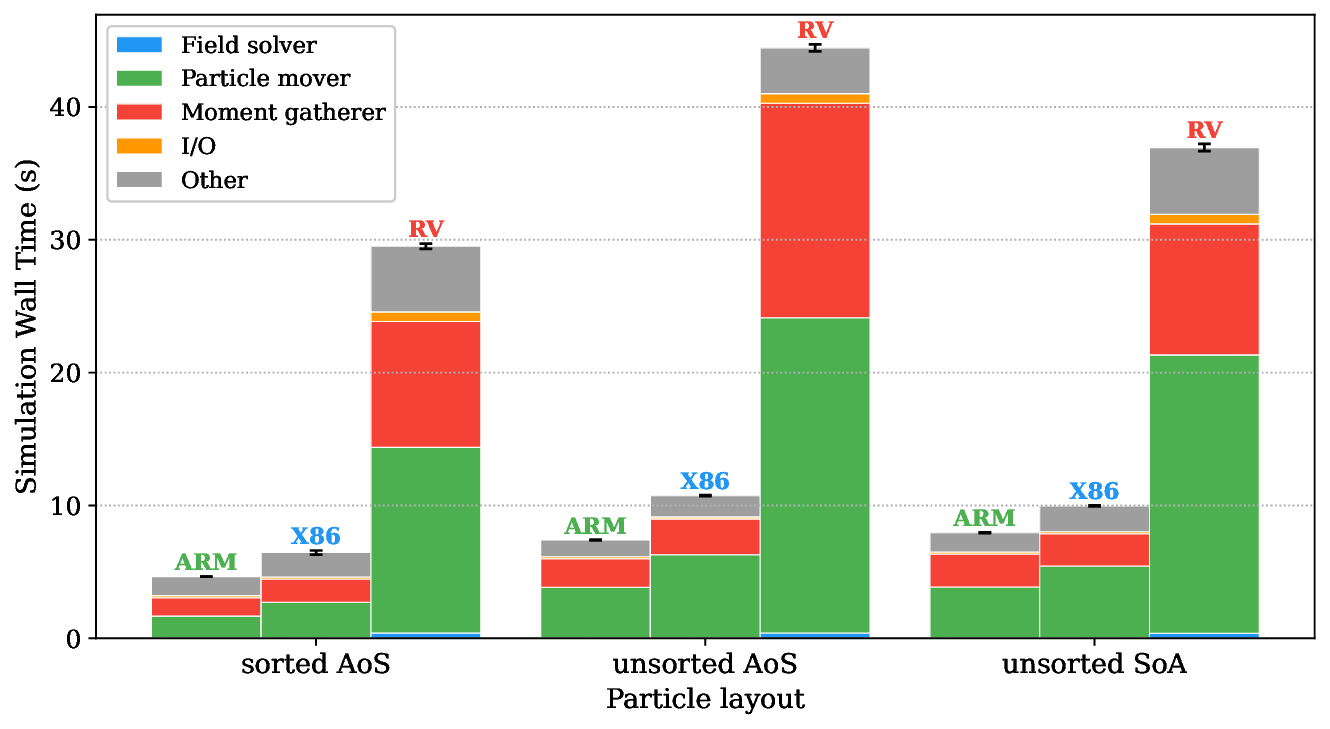}
\vspace{-0.5cm}
\caption{iPIC3D wall clock time breakdown by kernel phase across three particle memory layouts (sorted AoS, unsorted AoS, unsorted SoA) on ARM, x86, and RISC-V at 64 MPI ranks.}
\label{fig:ipic_walltimes_grouped}
\end{figure}


In Fig.~\ref{fig:ipic_ss_pe}, the strong-scaling wall time and the parallel efficiency are shown. The super-linear parallel efficiency observed on ARM and x86 at higher rank counts is attributed to improved  cache locality~\cite{ristov2016superlinear}: as the per-rank particle population and local field mesh shrink with increasing MPI ranks, the working set fits within the L2/L3 cache hierarchy, reducing memory latency for the irregular, position-dependent accesses inherent in particle–grid interpolation.  The pure MPI results span the full range of core counts available on the RISC-V node powered by the SG2044 CPU. Across all MPI ranks tested, the RISC-V platform remains consistently slower than both reference architectures. The slowdown relative to x86 (AMD EPYC 9554) is approximately $4.99\times$ (range: $4.84\times$--$5.24\times$), while relative to ARM Neoverse-V2 (NVIDIA GH200) is approximately $6.79\times$ (range: $6.23\times$--$7.52\times$). This behavior is consistent with the compiler analysis. As summarized in Table~\ref{ipic3d-compiler-summary}, loop-level SIMD vectorization is achieved in the particle mover and partially in the sorted moment accumulation kernels on ARM and x86, whereas the latter kernels remain largely non-vectorized on RISC-V. This leads to a reduced effective SIMD utilization on the RISC-V platform.

\begin{table}[t]
\caption{Comparison of vectorization efficiency for iPIC3D kernels across RISC-V (RVV), ARM (SVE), and x86 (AVX2). ``BB'' denotes basic-block partial vectorization (sub-loop); ``$\Delta$\,dp'' denotes the data-parallelism delta of RISC-V relative to ISA baseline.}
\vspace{-0.2cm}
\label{ipic3d-compiler-summary}
\scriptsize
\centering
\renewcommand{\arraystretch}{1.0}

\begin{tabular}{p{4.0cm} p{2.6cm} p{2.6cm} p{2.6cm} p{2.6cm}}
\toprule
\rowcolor{gray!15}
\textbf{Kernel} & \textbf{RISC-V (RVV)} & \textbf{x86 (AVX2)} & \textbf{ARM (SVE)} & \textbf{$\Delta$ dp (RISC-V vs. ARM/x86)} \\
\midrule

\multicolumn{5}{l}{\textit{Particle mover (hot path)}} \\
\addlinespace[2pt]
\texttt{mover\_PC\_AoS} 
  & \cellcolor{red!15}no vectorization 
  & \cellcolor{red!15}no vectorization 
  & \cellcolor{red!15}no vectorization 
  & — \\

\texttt{mover\_PC} (SoA) 
  & \cellcolor{yellow!20}1 partial BB (16\,B) 
  & \cellcolor{yellow!20}3 partial BBs (32\,B) 
  & \cellcolor{yellow!20}9 partial BBs (16\,B) 
  & \textcolor{red}{fewer BBs} \\

\midrule
\multicolumn{5}{l}{\textit{Moment gatherer (critical scatter kernel)}} \\
\addlinespace[2pt]
\texttt{sumMoments\_AoS} (unsorted) 
  & \cellcolor{red!15}no vectorization 
  & \cellcolor{red!15}no vectorization 
  & \cellcolor{red!15}no vectorization 
  & — \\

\texttt{sumMoments\_vec\_AoS} (sorted) 
  & \cellcolor{red!15}no vectorization 
  & \cellcolor{yellow!20}$\sim$4 partial BBs 
  & \cellcolor{yellow!20}$\sim$4 partial BBs 
  & \textcolor{red}{loss of BB} \\

\texttt{sumMoments\_vec} SoA (sorted) 
  & \cellcolor{red!15}no vectorization 
  & \cellcolor{yellow!20}$\sim$7 partial BBs 
  & \cellcolor{yellow!20}$\sim$10 partial BBs 
  & \textcolor{red}{loss of BB} \\

\midrule
\multicolumn{5}{l}{\textit{Data layout conversion and sorting}} \\
\addlinespace[2pt]
AoS $\leftrightarrow$ SoA / sorting 
  & \cellcolor{red!15}no vectorization 
  & \cellcolor{red!15}no vectorization 
  & \cellcolor{red!15}no vectorization 
  & — \\

\midrule
\multicolumn{5}{l}{\textit{Grid / solver utilities (non-critical path)}} \\
\addlinespace[2pt]

\texttt{dot} / \texttt{norm} / BLAS-like 
  & \cellcolor{green!20}RVV VL (scalable) 
  & \cellcolor{green!20}AVX2 32\,B 
  & \cellcolor{green!20}SVE scalable + 16\,B
  & \textcolor{green!60!black}{$\sim$ } \\

grid stencils (grad, Lapl.) 
  & \cellcolor{green!20}RVV VL (scalable) 
  & \cellcolor{green!20}AVX2 32\,B 
  & \cellcolor{green!20}SVE scalable + 16\,B
  & \textcolor{green!60!black}{$\sim$ } \\

\bottomrule
\end{tabular}
\end{table}

In Fig.~\ref{fig:ipic_walltimes_grouped}, we compare the end-to-end time across three particle memory layouts at MPI=64: sorted AoS (AoS with particles grouped by grid cell), unsorted AoS (baseline layout) and unsorted SoA. Particle sorting improves spatial locality by grouping particles belonging to the same cell. This transformation regularizes memory access patterns in the inner loops. On ARM and x86, this enables vectorization of the innermost accumulation kernel, as indicated in Table~\ref{ipic3d-compiler-summary}. However, on RISC-V, no loop-level vectorization is generated even in the sorted case, and the benefit is therefore attributed primarily to improved cache locality rather than SIMD execution. Given identical mathematical libraries and compiler settings, this result highlights a gap in compiler auto-vectorization maturity; future work will focus on identifying its root causes.

On RISC-V, the sorted AoS configuration reduces the total execution time by $1.51\times$ relative to the unsorted AoS ($29.52\pm0.20$ s versus $44.45 \pm 0.27$ s) and by $1.25\times$ relative to the unsorted SoA ($36.94 \pm 0.27$ s). The two unsorted layouts do not exhibit identical performance. The unsorted SoA configuration is $1.20\times$ faster than the unsorted AoS, whereas near-identical behavior is observed on ARM and x86. Since neither configuration is vectorized in the unsorted case, this difference is attributed to memory layout effects. The SoA layout provides contiguous access to particle attributes, which improves cache efficiency despite the absence of SIMD execution.


The moment accumulation kernel highlights this effect. The unsorted SoA configuration ($9.86 \pm 0.17$ s) is close to the sorted AoS case ($9.48 \pm 0.06$ s) and outperforms unsorted AoS ($16.15 \pm 0.07$ s) by $1.70\times$. Compiler reports confirm that all unsorted variants remain non-vectorized due to scatter dependencies,  while the sorted configurations expose several vectorizable basic blocks on ARM and x86 and none on RISC-V. The observed performance improvement is therefore driven by increased spatial locality rather than SIMD utilization.

The particle mover exhibits a different sensitivity. This kernel benefits strongly from particle sorting on all architectures as a result of improved locality in field interpolation and accumulation. In RISC-V, the sorted AoS configuration achieves a speedup of $1.70\times$ over unsorted AoS ($23.74 \pm 0.07$ s versus $13.97 \pm 0.05$ s) and $1.50\times$ over unsorted SoA ($20.95 \pm 0.01$ s). However, in contrast to ARM and x86, where multiple basic blocks are vectorized (9 and 3 respectively), the RISC-V implementation achieves only a single vectorized basic block in this kernel, as summarized in Table~\ref{ipic3d-compiler-summary}. Consequently, the performance gains on RISC-V are primarily due to improved memory access patterns rather than effective utilization of the vector unit.


\subsection{PLUTO}
PLUTO was benchmarked using the Orszag-Tang vortex test case, a standard for evaluating numerical stability and shock interactions. 
The simulations were carried out on a periodic 3D Cartesian domain using double-precision arithmetic. The numerical scheme employed the WENOZ reconstruction method, the HLLD Riemann solver, and third-order Runge-Kutta (RK3) time integration. To maintain the divergence-free condition of the magnetic field ( $\nabla \cdot  \mathbf{B} = 0$), a constrained transport algorithm was applied. The computational domain was discretized on a $256\times256\times256$ three-dimensional Cartesian grid and advanced for 10 time steps.
This configuration demonstrated high numerical stability and physical consistency, as evidenced by the maintenance of strict symmetry between the maximum and minimum extrema of the velocity and magnetic field components ($|V_{\max}| \approx |V_{\min}|$) alongside a well-behaved evolution of the Mach number in all integration steps.

\begin{figure}[t]
\centering
\includegraphics[width=0.49\textwidth]{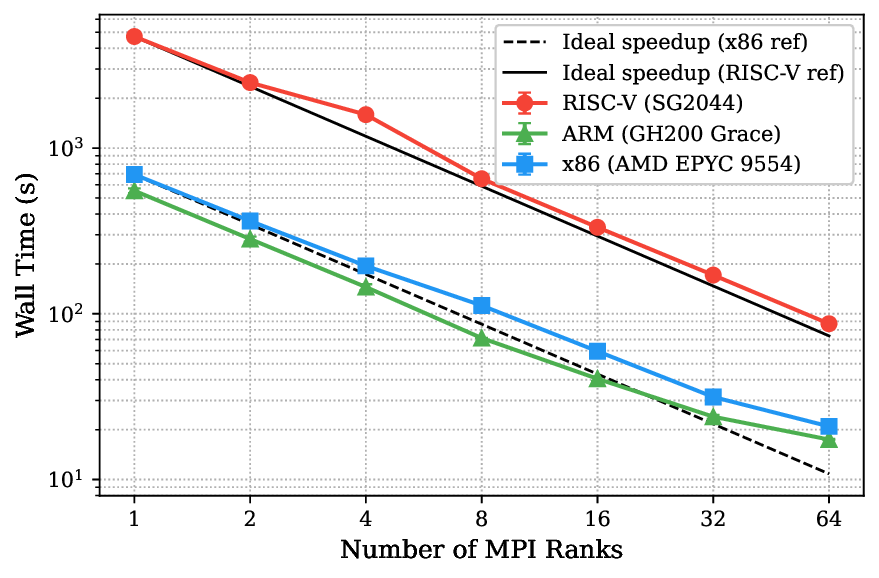}
\hfill
\includegraphics[width=0.49\textwidth]{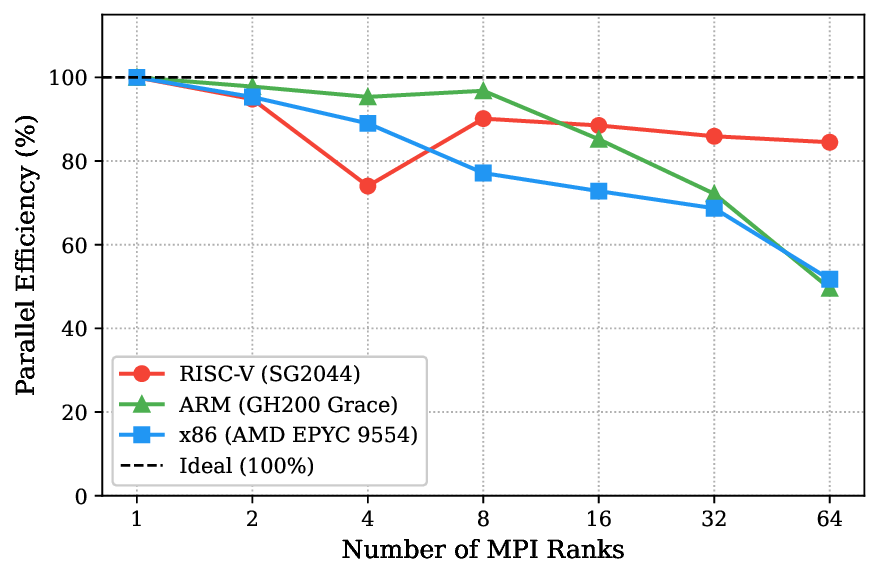}
\vspace{-0.5cm}
\caption{PLUTO strong-scaling results for the Orszag--Tang vortex test case on the same platforms of Fig.~\ref{fig:ipic_walltimes_grouped}. \textbf{Left:} wall clock time as a function of MPI rank count (lower is better). \textbf{Right:} parallel efficiency normalized to the single-rank execution time on each platform.}
\label{fig:gpluto_ss_pe}
\end{figure}

Strong-scaling wall clock time, parallel efficiency, and RISC-V slowdowns vs. x86 and ARM are presented in Fig.~\ref{fig:gpluto_ss_pe} and Table~\ref{pluto-slowdown}. As the MPI communicator size increases, end-to-end runtime decreases on every platform, confirming correct work distribution and functional MPI communication stacks on all architectures including RISC-V. Both ARM and x86 achieve ideal scaling, reflecting their mature memory subsystems and well-tuned MPI stacks. The RISC-V SG2044 also achieves a meaningful parallel speedup, demonstrating that the OpenMPI and GCC toolchain on the platform are functionally correct and capable of scaling the structured-grid MHD workload. At higher core counts, the RISC-V platform sustains better parallel efficiency above 80\%, whereas both ARM and x86 exhibit noticeable degradation, reaching approximately 50\% at 64 cores. This behavior is consistent with strong-scaling limitations arising from reduced per-rank workload. The comparatively lower computational throughput of RISC-V results in a more favorable balance between computation and communication, allowing it to maintain efficiency over a wider range of core counts.

\begin{table}[t]
\caption{PLUTO total simulation time and RISC-V slowdown relative to ARM and x86. Grid: $256^3$, 10 steps.}
\vspace{-0.2cm}
\label{pluto-slowdown}
\centering
\scriptsize
\begin{tabular*}{\linewidth}{@{\extracolsep{\fill}}
  c c c c c c c@{}}
\hline
\textbf{NP} &
\textbf{RISC-V (s)} &
\textbf{ARM (s)} &
\shortstack[c]{\textbf{Slowdown}\\\textbf{vs ARM}} &
\textbf{x86 (s)} &
\shortstack[c]{\textbf{Slowdown}\\\textbf{vs x86}} \\
\hline
 1 & $4711.5 \pm 7.0$ & $553.0 \pm 63.3$ & $8.52\times$ & $692.7 \pm 0.4$ & $6.80\times$ \\
 2 & $2484.9 \pm 5.0$ & $282.7 \pm 32.3$ & $8.79\times$ & $363.4 \pm 0.4$ & $6.84\times$ \\
 4 & $1591.3 \pm 6.7$ & $145.0 \pm 19.4$ & $10.97\times$ & $194.6 \pm 0.4$ & $8.18\times$ \\
 8 &  $653.3 \pm 2.6$ &  $71.4 \pm 0.2$ & $9.15\times$ & $112.3 \pm 0.4$ & $5.82\times$ \\
16 &  $332.7 \pm 0.8$ &  $40.6 \pm 4.4$ & $8.20\times$ &  $59.5 \pm 0.1$ & $5.60\times$ \\
32 &  $171.4 \pm 0.4$ &  $23.9 \pm 2.1$ & $7.16\times$ &  $31.5 \pm 0.0$ & $5.44\times$ \\
64 &   $87.1 \pm 0.2$ &  $17.4 \pm 0.5$ & $5.00\times$ &  $20.9 \pm 0.1$ & $4.16\times$ \\
\hline
\end{tabular*}
\end{table}

As shown in Table~\ref{pluto-slowdown}, RISC-V consistently exhibits the slowest runtimes across all tested MPI rank counts. Averaged across all MPI configurations, it shows a slowdown of $8.26\times$ compared to ARM (range: $5.00\times$--$10.97\times$) and $6.12\times$ compared to x86 (range: $4.16\times$--$8.18\times$).
The performance gap peaks at $NP=4$ (where $NP$ denotes the number of MPI ranks), reaching $10.97\times$ relative to ARM and $8.18\times$ relative to x86.
This behavior is primarily attributed to reduced single-core peak performance. The PLUTO MHD solver is dominated by stencil-based updates and Riemann solver evaluations, which involve intensive double-precision floating-point operations. These kernels are sensitive to the peak floating-point performance and the degree of data-level parallelism supported by the hardware.


\begin{table}[t]
\caption{Same as Table~\ref{ipic3d-compiler-summary}, but for the code PLUTO.}
\vspace{-0.2cm}
\label{compiler-summary}
\centering
\scriptsize
\setlength{\tabcolsep}{5pt}
\renewcommand{\arraystretch}{1.}

\begin{tabular}{p{4.6cm} p{2.4cm} p{2.5cm} p{2.4cm} p{1.3cm}}
\toprule
\rowcolor{gray!15}
\textbf{Kernel} & \textbf{RISC-V (RVV)} & \textbf{x86 (AVX-512)} & \textbf{ARM (SVE-128)} & \textbf{$\Delta$ dp} \\
\midrule

\multicolumn{5}{l}{\textit{Riemann solvers — innermost hot path}} \\

\texttt{roe.cpp} — main flux loop & \cellcolor{yellow!20}8 B : 1 dp : scalar & \cellcolor{green!20}64 B : 8 dp & \cellcolor{red!15}no vectorization & \textcolor{red}{8×↓} \\

\texttt{hlld.cpp} — HLLD solver body & \cellcolor{red!15}no vectorization & \cellcolor{blue!15}64 B partial block & \cellcolor{red!15}no vectorization & — \\

\texttt{hllc.cpp} — solver body & \cellcolor{red!15}no vectorization & \cellcolor{blue!15}64 B partial block & \cellcolor{red!15}no vectorization & — \\

\texttt{hll.cpp} + \texttt{tvdlf.cpp} + \texttt{rhs.cpp} & \cellcolor{red!15}no vectorization & \cellcolor{red!15}no vectorization & \cellcolor{red!15}no vectorization & — \\

\midrule
\multicolumn{5}{l}{\textit{Reconstruction}} \\

\texttt{array\_reconstruct.cpp} & \cellcolor{green!20}128 B : 16 dp & \cellcolor{green!10}64 B : 8 dp & \cellcolor{blue!10}16 B : 2 dp & \textcolor{green!50!black}{2×↑} \\

\texttt{mp5\_states.cpp} & \cellcolor{blue!10}16 B : 2 dp : 1 site & \cellcolor{blue!15}64 B : 8 dp : 2 sites & \cellcolor{red!15}no vectorization & \textcolor{red}{4×↓} \\

\texttt{mhd.hpp} — PrimEigenvectors loop & \cellcolor{green!20}128 B : 16 dp & \cellcolor{red!15}no vectorization & \cellcolor{blue!10}16 B : 2 dp & —  \\

\texttt{mhd.hpp} — inlined bb partials & \cellcolor{blue!10}16 B : 2 dp & \cellcolor{blue!15}64 B : 8 dp & \cellcolor{blue!10}16 B : 2 dp & \textcolor{red}{4×↓} \\

\midrule
\multicolumn{5}{l}{\textit{Stencil update / time integration}} \\

\texttt{update\_stage.cpp} & \cellcolor{green!20}128 B : 16 dp & \cellcolor{green!10}64 B : 8 dp & \cellcolor{blue!10}16 B : 2 dp & \textcolor{green!50!black}{2×↑} \\

\texttt{rk\_step.cpp} & \cellcolor{green!20}128 B : 16 dp & \cellcolor{green!10}64 / 32 / 16 B & \cellcolor{blue!10}16 B : 2 dp & \textcolor{green!50!black}{2×↑} \\

\midrule
\multicolumn{5}{l}{\textit{Constrained transport}} \\

\texttt{ct\_emf\_average.cpp} & \cellcolor{green!20}128 / 16 B & \cellcolor{green!10}64 / 32 / 16 B & \cellcolor{blue!10}16 / 8 B & $\sim$ \\

\texttt{ct\_update.cpp} & \cellcolor{green!20}128 B : 16 dp & \cellcolor{green!10}64 B : 8 dp & \cellcolor{blue!10}16 B : 2 dp & \textcolor{green!50!black}{2×↑} \\

\bottomrule
\end{tabular}
\end{table}

The performance difference is primarily driven by a combination of hardware limitations and vectorization compiler support on RISC-V hardware. A comparative analysis of RISC-V and x86, based on compiler log inspection, is reported in Table~\ref{compiler-summary}. The RISC-V platform, when configured with RVV LMUL=8, can process up to 16 double-precision elements per instruction by grouping multiple vector registers.
This represents a two-fold increase in per-instruction vector width compared to x86 AVX-512, which operates on 64-byte registers. In simple alias-free stencil loops, such as the conservative update stage and the Runge-Kutta time-stepping kernel, this architectural capability is effectively utilized. Under these conditions, up to 16 double-precision elements are processed per instruction on RISC-V, whereas only 8 elements are processed per instruction on x86. Despite the advantage, the overall system performance is approximately $6.12\times$ slower. This discrepancy arises because the hardware benefit is confined to kernels that are not performance critical. In MHD simulations, the dominant computational cost is associated with the Riemann solvers. In these regions, the RISC-V compiler backend exhibits limited effectiveness, resulting in poor vectorization and computational throughput. 
For example, the main flux loop of the Roe solver (\texttt{roe.cpp}) is executed for every cell face, in every sweep direction at every Runge-Kutta sub-stage. On RISC-V, this loop is compiled to an 8-byte vector width, which means that only one double-precision element is processed per instruction. This is effectively the same as scalar execution. On x86, the same loop is vectorized to 64 bytes, allowing 8 elements per instruction. This creates an \textbf{$8\times$} throughput difference in the most frequently executed loop. No vectorization is observed on ARM for this kernel, which has the same vector length as RISC-V; the reason is under investigation. The HLLD and HLLC solvers are not vectorized at all on RISC-V, while x86 achieves partial vectorization with 64-byte blocks. Other kernels, such as HLL, TVDLF, and the right-hand-side flux differencing, remain unvectorized on both platforms.

A similar limitation is observed in the higher-order reconstruction stage. The MP5/WENOZ stencil (\texttt{mp5\_states.cpp}) reaches only 16-byte vectorization on RISC-V (two double-precision elements), whereas x86 achieves 64-byte vectorization at selected sites. Likewise, the inlined eigenvector routines in \texttt{mhd.hpp}, which are instantiated throughout the codebase, are only partially vectorized at 16 bytes on RISC-V, compared to 64 bytes on x86. This results in an effective \textbf{$4\times$} reduction in per-instruction throughput in these components.


Another source of disparities in auto-vectorization runtime impact is accountable to the compiler own alias analysis.
On RISC-V with the GCC~14.2 toolchain, 380 out of 427 auto-vectorized loops (89\%) are correctly multi-versioned and guarded by runtime aliasing checks.
We observe similar code generation on x86 with GCC~12.4 for 331 out of 462 (72\%) auto-vectorized loops.
Given the higher fraction of alias-guarded loops on RISC-V (89\% vs. 72\% on x86), runtime scalar fallbacks are expected to occur more frequently on RISC-V, further limiting the effective utilization of its wider vector units.
Compared to ARM, RISC-V exhibits broadly similar auto-vectorization behavior, suggesting that the observed performance differences may partly stem from hardware maturity.

Memory subsystem characteristics contribute a secondary but non-negligible share of the slowdown. PLUTO stores multiple conserved and primitive variables per grid cell, and stencil operations require repeated traversal of these arrays across neighboring cells, generating substantial streaming memory traffic. Performance is further constrained by memory‑system delays when the RISC‑V cache hierarchy exhibits lower bandwidth or higher latency than the reference architectures. At higher MPI ranks, the local subdomain handled by each process becomes smaller, and inter-process communication consumes a larger fraction of the total runtime. As a result, computational inefficiency is partially masked, leading to a smaller observed relative slowdown, consistent with the decrease from $8.18\times$ at $NP = 4$ to $4.16 \times$ at $NP = 64$ compared to x86.

\subsection{\OG~}

\begin{figure}[t]
\centering
\includegraphics[width=\textwidth]{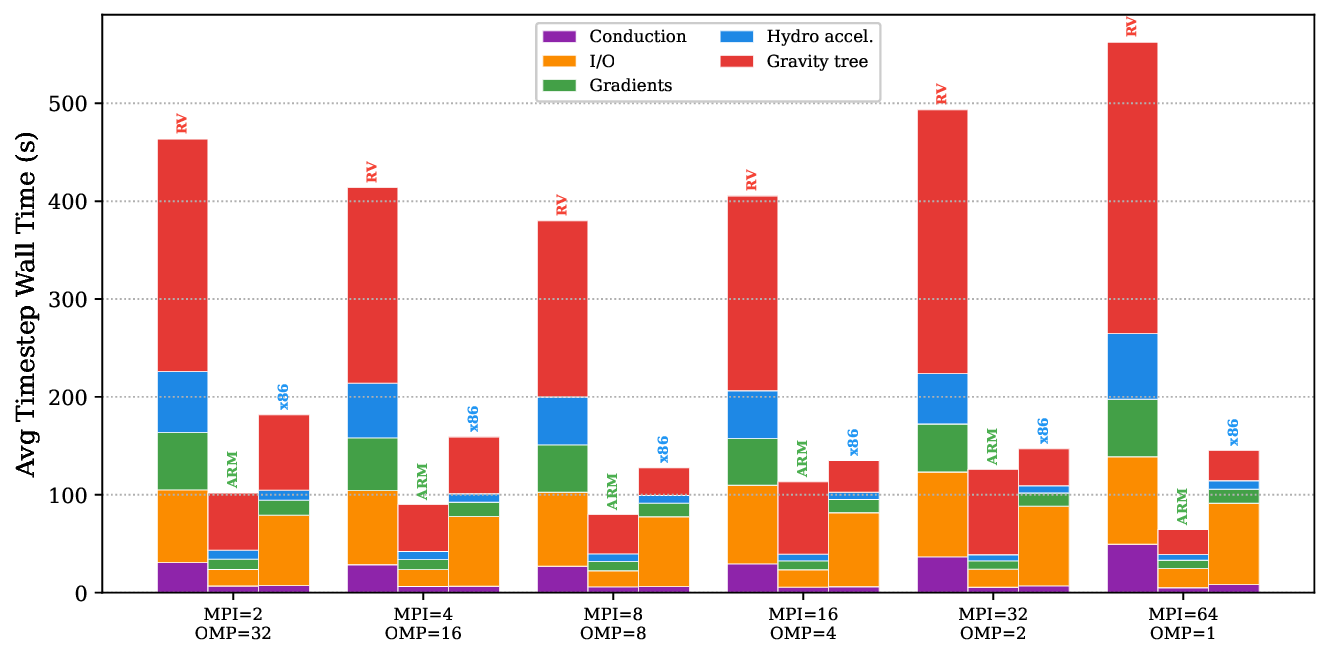}
\vspace{-0.5cm}
\caption{\OG~ per-kernel wall clock time breakdown for the production cosmological test case on RISC-V, ARM, and x86 across various MPI+OpenMP configurations over four steps across 10 runs.
}
\label{fig:og_kernel}
\end{figure}

The performance of \OG~ was evaluated using a production-grade galaxy formation initial condition, focusing on the hybrid MPI+OpenMP scaling across the full 64-core range of the SG2044.
This approach allowed us to identify the optimal balance between inter-rank communication and intra-node memory bandwidth utilization.
To optimize locality, process binding was enforced via \texttt{--map-by <object>:PE=\textit{N}} with the \texttt{close} thread policy, reserving $N$ contiguous cores per rank aligned to the nearest shared hardware locality domain. On SG2044, ranks were bound to 4-core L2 cache clusters for low thread counts (OMP\,$\leq$\,4) and to sockets for higher counts; on GH200 and AMD EPYC, where L2 caches are private to each core, socket-level binding was used throughout.

In Fig.~\ref{fig:og_kernel}, the total wall time accumulated over four simulation steps is decomposed by kernel over the full range of hybrid configurations. RISC-V is consistently slower than ARM and x86 by a factor of 3.69--8.47$\times$ and 2.69--6.10$\times$, respectively, across all configurations including MPI=1, OMP=64. The relative slowdown of RISC-V versus ARM rises sharply at MPI=64, OMP=1, reaching 8.47$\times$ where ARM performs particularly well. The gravity tree is identified as the dominant kernel in all cases. For MPI $\geq$ 4, it accounts for approximately 43--50\% of the total wall time. 
In the single-rank configuration (MPI=1, OMP=64), the gravity tree alone contributes roughly 89\% of the total wall time. This trend is expected, as communication overhead is removed and parallelism must be extracted entirely through shared-memory threading. The hydrodynamic acceleration and gradient phases represent the next largest contributions in configurations that are more evenly balanced. Table~\ref{tab:og_hybrid} summarizes these data, providing a quantitative overview.

\begin{table}[t]
\caption{RISC-V total, gravity tree, and tree walk wall times, accumulated over four steps across 10 runs, with slowdown relative to the MPI\,=\,8, OMP\,=\,8 baseline ($\dagger$).}\label{tab:og_hybrid}
\centering
\scriptsize
\vspace{-0.2cm}
\begin{tabular*}{\linewidth}{@{\extracolsep{\fill}}
  >{\raggedright\arraybackslash}l
  c c c c c c c c@{}}
\hline
\centering
\textbf{Configuration} &
\textbf{Total (s)} &
\shortstack[c]{\textbf{Slow-}\\\textbf{down}} &
\shortstack[c]{\textbf{Gravity}\\\textbf{tree (s)}} &
\shortstack[c]{\textbf{Gravity}\\\textbf{slowdown}} &
\shortstack[c]{\textbf{Tree}\\\textbf{walk (s)}} &
\shortstack[c]{\textbf{Walk}\\\textbf{slowdown}} &
\shortstack[c]{\textbf{RV/}\\\textbf{ARM}} &
\shortstack[c]{\textbf{RV/}\\\textbf{x86}} \\
\hline
MPI=1,  OMP=64 & $8935.0 \pm 3787.5$ & $20.73\times$ & $7973.3 \pm 3469.2$ & $44.30\times$ & $7949.8 \pm 3459.7$ & $46.90\times$ & $6.03\times$ & $6.10\times$ \\
MPI=2,  OMP=32 &  $549.6 \pm   2.1$ &  $1.28\times$ &  $237.4 \pm   1.2$ &  $1.32\times$ &  $224.2 \pm   1.9$ &  $1.32\times$ & $4.64\times$ & $2.69\times$ \\
MPI=4,  OMP=16 &  $480.2 \pm   2.6$ &  $1.11\times$ &  $199.8 \pm   1.1$ &  $1.11\times$ &  $188.3 \pm   1.3$ &  $1.11\times$ & $4.66\times$ & $2.74\times$ \\
MPI=8,  OMP=8  &  $430.9 \pm   1.7$ & $1.00\times^\dagger$ & $180.0 \pm   0.7$ & $1.00\times^\dagger$ & $169.5 \pm   1.3$ & $1.00\times^\dagger$ & $4.76\times$ & $3.07\times$ \\
MPI=16, OMP=4  &  $452.2 \pm   5.7$ &  $1.05\times$ &  $199.0 \pm   4.5$ &  $1.11\times$ &  $179.0 \pm   2.6$ &  $1.06\times$ & $3.69\times$ & $3.07\times$ \\
MPI=32, OMP=2  &  $536.9 \pm   6.6$ &  $1.25\times$ &  $269.6 \pm   6.9$ &  $1.50\times$ &  $239.0 \pm   9.3$ &  $1.41\times$ & $4.00\times$ & $3.35\times$ \\
MPI=64, OMP=1  &  $613.0 \pm   5.4$ &  $1.42\times$ &  $297.8 \pm   4.7$ &  $1.65\times$ &  $274.4 \pm   7.1$ &  $1.62\times$ & $8.47\times$ & $3.87\times$ \\
\hline
\end{tabular*}
\end{table}

The fastest configurations are observed for MPI=8, OMP=8 and MPI=16, OMP=4 with total wall times of $430.9 \pm 1.6$ s and $452.2 \pm 5.4$ s, respectively. This behavior is consistent with the hardware topology of the SG2044 processor, where four cores share a 2 MiB L2 cache cluster. It should be noted that thread count does not directly depend on the L2 cache size. Instead, performance is influenced by how the thread count aligns with the cache-sharing topology. When threads are scheduled on cores that share an L2 cache, data locality is improved and cache coherency overhead is reduced. When the OpenMP thread count is constrained to match this hardware boundary, data reuse during the tree-walk phase is increased. At the same time, inter-cluster cache coherency traffic is minimized. As a result, the lowest wall times are observed for these configurations.


The performance breakdown in Fig.~\ref{fig:og_kernel} reveals distinct architectural sensitivities.
The gravity tree and hydrodynamic acceleration phases are memory-bound and involve irregular, indirect pointer-chasing traversals over particle data. Their working sets exceed L3 cache capacity, leading to non-sequential DRAM access. The gravity tree walk is not vectorized on any platform. Its inner loop contains a data-dependent branch (the Barnes–Hut opening criterion) and indirect pointer accesses, both of which prevent SIMD execution. This is confirmed by compiler diagnostics reporting unsupported control flow and complex access patterns. A similar limitation is observed in the SPH force kernel, where neighbor interactions require indirect indexing (e.g., \texttt{P[ngblist[i]]}) resulting in gather-like access over AoS data. These limitations arise from the algorithmic structure rather than the compiler. Most non-vectorized loops are therefore associated with control flow and irregular memory access, while vectorized loops are confined to auxiliary tasks outside the critical execution path.

The RISC-V memory subsystem further amplifies the memory-bound nature of these kernels. The SG2044 employs DDR5-4266 memory, which provides lower per-channel bandwidth and lower aggregate throughput compared to the DDR5-4800 subsystem of the AMD EPYC 9554, which also offers a higher number of memory channels~\cite{Diehl2024_RISCV,Brown2023_RAJAPerf,Brown2024_SG2042,Brown2025_SG2044}. As a result, cache-miss requests contend for shared DRAM resources and are served in a queued manner. This reduces effective memory-level parallelism, increases access latency, and limits achievable bandwidth. For partially vectorized kernels, such as the MFM gradient computation and small dense matrix operations, performance is additionally constrained by the vector unit. The C920v2 implements RVV 1.0 with VLEN=128 bits, corresponding to two double-precision elements per instruction. This represents a fourfold reduction in per-instruction throughput compared to AVX-512 on x86, which processes eight double-precision elements. For the gravity tree and SPH kernels, which remain scalar on all platforms, this difference is not relevant. However, it directly limits throughput in those kernels that do vectorize (see Table~\ref{tab:og3-compiler-summary} for a summary of the vectorization efficiency). The performance gap is further influenced by lower memory-level parallelism and reduced latency hiding. The out-of-order execution window and hardware prefetching mechanisms of the C920v2 are less aggressive than those of AMD Zen\,4 and ARM Neoverse,V2, which provide deeper reorder buffers, more outstanding memory requests, and more effective prefetching strategies~\cite{Bramas2023_SPC5}. 

\begin{table}[t]
\caption{Comparison of vectorization efficiency across RISC-V (RVV), ARM (SVE), and x86 (AVX-512) for \OG~. BB is basic-block vectorization only (no full loop) and $\Delta$\,dp is RISC-V relative to x86.
}
\vspace{-0.2cm}
\label{tab:og3-compiler-summary}
\centering
\scriptsize
\setlength{\tabcolsep}{5pt}
\renewcommand{\arraystretch}{1.15}

\begin{tabular}{p{5.5cm} p{2.5cm} p{2.5cm} p{2.5cm} p{0.8cm}}
\toprule
\rowcolor{gray!15}
\textbf{Kernel} & \textbf{RISC-V (RVV)} & \textbf{ARM (SVE)} & \textbf{x86 (AVX-512)} & \textbf{$\Delta$ dp} \\
\midrule

\multicolumn{5}{l}{\textit{N-body gravity}} \\

Gravity/\texttt{gravtree.cpp} — tree walk (traversal)
  & \cellcolor{red!15}no vectorization
  & \cellcolor{red!15}no vectorization
  & \cellcolor{red!15}no vectorization
  & — \\

Gravity/\texttt{forcetree.cpp} — tree build (aux. loops)
  & \cellcolor{yellow!20}2–16\,B + 8–16\,B BB
  & \cellcolor{yellow!25}8–16\,B + 16\,B BB
  & \cellcolor{green!10}32–64\,B : 4–8\,dp + 64\,B BB
  & \textcolor{red}{2–4×↓} \\

Gravity/\texttt{gravtree.cpp} — aux. loops (BB)
  & \cellcolor{yellow!20}8–16\,B BB
  & \cellcolor{yellow!25}16\,B BB
  & \cellcolor{blue!15}64\,B BB
  & \textcolor{red}{4–8×↓} \\

\midrule
\multicolumn{5}{l}{\textit{Hydrodynamics — SPH/MFM}} \\

Hydro/\texttt{hydra.cpp} — SPH force inner loop
  & \cellcolor{red!15}no vectorization
  & \cellcolor{red!15}no vectorization
  & \cellcolor{red!15}no vectorization
  & — \\

Hydro/\texttt{hydra.cpp} — auxiliary (BB)
  & \cellcolor{yellow!25}16\,B BB
  & \cellcolor{yellow!25}16\,B BB
  & \cellcolor{blue!15}64\,B BB
  & \textcolor{red}{4×↓} \\

\midrule
\multicolumn{5}{l}{\textit{Cooling \& chemistry}} \\

CoolingSfr/\texttt{cooling.cpp} — ODE integration loop
  & \cellcolor{red!15}no vectorization
  & \cellcolor{red!15}no vectorization
  & \cellcolor{red!15}no vectorization
  & — \\

CoolingSfr/\texttt{cooling.cpp} — table lookup (BB)
  & \cellcolor{yellow!20}2–16\,B BB
  & \cellcolor{yellow!25}16\,B BB
  & \cellcolor{blue!15}64\,B BB
  & \textcolor{red}{4–32×↓} \\

Hydro/\texttt{compute\_gradients.cpp} — gradient loops
  & \cellcolor{yellow!20}2–16\,B : 0.25–2\,dp (versioned)
  & \cellcolor{green!15}16\,B + VLV (versioned)
  & \cellcolor{yellow!25}8–16\,B : 1–2\,dp (versioned)
  & $\sim$ \\

\midrule
\multicolumn{5}{l}{\textit{Halo finding — FOF/Subfind}} \\

FofSubfind/\texttt{subfind\_collective.cpp} — group accumulation
  & \cellcolor{red!15}no vectorization
  & \cellcolor{red!15}no vectorization
  & \cellcolor{red!15}no vectorization
  & — \\

FofSubfind/\texttt{subfind\_findlinkngb.cpp} — linking walk
  & \cellcolor{red!15}no vectorization
  & \cellcolor{red!15}no vectorization
  & \cellcolor{red!15}no vectorization
  & — \\

FofSubfind/\texttt{subfind\_findlinkngb.cpp} — irregular BB
  & \cellcolor{yellow!20}8\,B : 1\,dp BB
  & \cellcolor{yellow!25}16\,B BB
  & \cellcolor{blue!15}64\,B BB
  & \textcolor{red}{8×↓} \\

\midrule
\multicolumn{5}{l}{\textit{Particle data operations}} \\

CoolingSfr/\texttt{rearrange\_particles.cpp} — particle swap
  & \cellcolor{blue!15}16\,B BB only
  & \cellcolor{blue!15}16\,B BB only
  & \cellcolor{blue!15}64\,B BB only
  & \textcolor{red}{4×↓} \\

\midrule
\multicolumn{5}{l}{\textit{Domain decomposition \& sort}} \\

CodeBase/\texttt{global.cpp} — cost accumulation loops
  & \cellcolor{yellow!25}16\,B : 2\,dp
  & \cellcolor{yellow!25}16\,B : 2\,dp
  & \cellcolor{green!10}32\,B : 4\,dp
  & \textcolor{red}{2×↓} \\

CodeBase/\texttt{global.cpp} — reduction BB
  & \cellcolor{yellow!25}16\,B BB
  & \cellcolor{yellow!25}16\,B BB
  & \cellcolor{blue!15}64\,B BB
  & \textcolor{red}{4×↓} \\

PM/\texttt{parallel\_sort\_l3.cpp} — merge copy
  & \cellcolor{yellow!25}16\,B : 2\,dp (versioned)
  & \cellcolor{yellow!25}16\,B : 2\,dp + VLV
  & \cellcolor{green!20}64\,B : 8\,dp (versioned)
  & \textcolor{red}{4×↓} \\

\midrule
\multicolumn{5}{l}{\textit{Time integration}} \\

Integrator/\texttt{timestep.cpp} — timestep loop
  & \cellcolor{yellow!25}16\,B : 2\,dp + 8\,B BB
  & \cellcolor{red!15}no loop vec; 16\,B BB only
  & \cellcolor{green!10}32\,B : 4\,dp + 64\,B BB
  & \textcolor{red}{2×↓} \\

System/\texttt{system.cpp} — integrator reductions
  & \cellcolor{yellow!25}16\,B : 2\,dp
  & \cellcolor{yellow!25}16\,B : 2\,dp + VLV
  & \cellcolor{green!20}64\,B : 8\,dp (versioned)
  & \textcolor{red}{4×↓} \\

System/\texttt{system.cpp} — narrow scalar residual
  & \cellcolor{red!20}2\,B : $\sim$scalar
  & \cellcolor{yellow!25}16\,B VLV
  & \cellcolor{green!20}64\,B : 8\,dp
  & \textcolor{red}{32×↓} \\








\bottomrule
\end{tabular}

\medskip
\noindent\scriptsize
The corresponding raw compiler summary data are available upon request for reproducibility purposes.
\end{table}

A significant performance degradation is observed as the decomposition shifts toward fewer MPI ranks and higher OpenMP thread counts. A slowdown of up to $20.7\times$ is measured at MPI=1, OMP=64 relative to the MPI=8, OMP=8 configuration. This behavior arises from two related effects. First, as the number of threads increases, the combined working set of memory-bound kernels, particularly the gravity tree walk, exceeds the 2~MiB L2 cache shared within each core cluster. This leads to frequent cache evictions and increased reliance on the shared L3 cache and DRAM. Second, memory bandwidth is saturated at relatively low thread counts. Beyond this point, performance is limited by data movement rather than computation. This saturation behavior is consistent with memory-bound workloads, where throughput plateaus once a moderate number of cores is reached. Thread placement cannot mitigate this limitation once the cache capacity and memory bandwidth are exhausted.

At lower thread counts, intra-node memory contention is reduced. However, increasing the number of MPI ranks introduces communication overhead. This includes particle exchange across domain boundaries and collective operations such as global energy diagnostics. These costs offset the benefits of reduced memory pressure and lead to higher wall times at large MPI counts. The configurations MPI=8, OMP=8 and MPI=16, OMP=4 represent a balance between memory bandwidth contention and communication overhead. These points correspond to the minimum observed wall time and are consistent with the underlying hardware and communication characteristics.

\section{Summary and Future Directions}
\label{summary}
This work presents an end-to-end evaluation of production astrophysical applications on a modern RISC-V platform, demonstrating application-level portability without changes to numerical algorithms. However, a clear performance gap remains, with RISC-V being about $3-9\times$ slower relative to the x86 and ARM reference platforms, depending on the workload. The gap is mainly attributed to limitations in memory bandwidth, cache hierarchy, vector width, and less-mature auto-vectorization capability of the GNU compiler suite. In particular, the 128-bit vector unit and lower instruction throughput reduce compute performance, while memory-bound and irregular kernels are further affected by early memory bandwidth saturation, leading to reduced scalability at higher thread counts.

The study identifies key architecture-specific bottlenecks and shows that performance is strongly influenced by workload characteristics and hybrid MPI + OpenMP configurations. Targeted optimizations, especially those that improve memory locality, provide measurable gains while preserving portability. These results show that RISC-V can support complex scientific workloads and highlight the main areas where improvements are required to achieve competitive performance. As a future direction, a single kernel will be isolated for detailed analysis and optimization, with a focus on adaptation to RISC-V. Performance improvements are expected from enhanced explicit-vectorization, data layout restructuring, and more efficient memory access patterns. In particular, approaches that improve spatial locality and reduce irregular memory access are likely to mitigate current hardware limitations.

Although this study focuses primarily on software portability and efficiency, the observed performance limitations also point to the necessary architectural advances for RISC-V to achieve competitiveness in the HPC domain. In particular, future designs should prioritize higher-bandwidth memory subsystems, such as the integration of HBM, and more efficient cache hierarchies to mitigate the saturation effects observed in memory-bound astrophysical kernels. Additionally, increasing the physical width of vector units, together with continued improvements in compiler auto-vectorization capabilities, is likely to be essential for reducing the performance gap with established x86 and ARM platforms.

\begin{credits}
\subsubsection{\ackname} 
This work benefited from support provided by the SPACE project, funded by the European Union. This project has received funding
from the European High Performance Computing Joint Undertaking (JU) and from Belgium, the Czech Republic, France,
Germany, Greece, Italy, Norway, and Spain under grant agreement No. 101093441.
Authors gratefully acknowledge access to the E4 Computer Engineering datacentre (Scandiano, Italy) and the Monte Cimone RISC-V Cluster at Università degli Studi di Bologna (Bologna, Italy).


\end{credits}
%
\bibliographystyle{splncs04}
\bibliography{paper}

@inproceedings{Shukla2025_SPACE_CoE,
  author    = {Shukla, N and others},
  title     = {EuroHPC SPACE CoE: Redesigning Scalable Parallel Astrophysical Codes for Exascale (Invited Paper)},
  booktitle = {Proceedings of the 22nd ACM International Conference on Computing Frontiers Workshops and Special Sessions},
  series    = {CF '25 Companion},
  pages     = {177--184},
  publisher = {ACM},
  year      = {2025},
  month     = jul,
  doi       = {10.1145/3706594.3728892}
}

@article{Shukla2025_EuroHPCDay,
  author  = {Shukla, N and others},
  title   = {Towards Exascale Computing for Astrophysical Simulation Leveraging the Leonardo EuroHPC System},
  journal = {Procedia Computer Science},
  year    = {2025},
  pages   = {112--123},
  doi     = {10.1016/j.procs.2025.08.238267}
}

@article{Shukla2026_exascale,
  author  = {Shukla, N. and others},
  title   = {Exascale Computing to Accelerate Discoveries in Astrophysics and Space Plasma Physics},
  journal = {Nature Astronomy},
  year    = {2026},
  doi     = {10.1038/s41550-026-02807-8}
}

@incollection{Venieri2026,
  author    = {Venieri, Emanuele and Manoni, Simone and others},
  title     = {Monte Cimone v2: HPC RISC-V Cluster Evaluation and Optimization},
  booktitle = {High Performance Computing},
  editor    = {Neuwirth, Sarah and others},
  series    = {Lecture Notes in Computer Science},
  volume    = {16091},
  pages     = {576--585},
  publisher = {Springer},
  address   = {Cham},
  year      = {2026},
  doi       = {10.1007/978-3-032-07612-0_44}
}

@article{BarnesHut1986,
  author  = {Barnes, J and Hut, P},
  title   = {A Hierarchical O(N log N) Force-Calculation Algorithm},
  journal = {Nature},
  year    = {1986},
  volume  = {324},
  number  = {6096},
  pages   = {446--449},
  doi     = {10.1038/324446a0}
}

@article{Groth2023,
  author  = {Groth, F and others},
  title   = {The Cosmological Simulation Code OpenGadget3: Implementation of Meshless Finite Mass},
  journal = {Monthly Notices of the Royal Astronomical Society},
  year    = {2023},
  volume  = {526},
  number  = {1},
  pages   = {616--644},
  doi     = {10.1093/mnras/stad2717}
}

@article{Diehl2024_RISCV,
  author  = {Diehl, P  and others},
  title   = {Preparing for HPC on RISC-V: Examining Vectorization and Distributed Performance of an Astrophysics Application with HPX and Kokkos},
  journal = {arXiv preprint arXiv:2407.00026},
  year    = {2024},
  doi     = {10.48550/arXiv.2407.00026}
}

@article{Diehl2023_OctoTiger,
  author  = {Diehl, P and Dai{\ss}, G and others},
  title   = {Evaluating HPX and Kokkos on RISC-V Using an Astrophysics Application (Octo-Tiger)},
  journal = {arXiv preprint arXiv:2309.06530},
  year    = {2023}
}

@article{Bramas2023_SPC5,
  author  = {Bramas, B{\'e}renger},
  title   = {SPC5: An Efficient SpMV Framework Vectorized Using ARM SVE and x86 AVX-512},
  journal = {arXiv preprint arXiv:2307.14774},
  year    = {2023}
}

@inproceedings{Brown2024_SG2042,
  author    = {Brown, N and Jamieson, M},
  title     = {Performance Characterisation of the 64-Core SG2042 RISC-V CPU for HPC},
  booktitle = {ISC High Performance 2024 Workshops},
  series    = {Lecture Notes in Computer Science},
  volume    = {15058},
  publisher = {Springer},
  year      = {2025}
}

@inproceedings{Brown2023_RAJAPerf,
  author    = {Brown, N and others},
  title     = {Initial RAJAPerf Evaluation on Sophon SG2042},
  booktitle = {Proceedings of the SC'23 Workshops},
  year      = {2023}
}

@article{Brown2025_SG2044,
  author  = {Brown, N and Jamieson, M},
  title   = {Is RISC-V Ready for High Performance Computing? An Evaluation of the Sophon SG2044},
  journal = {arXiv preprint arXiv:2508.13840},
  year    = {2025}
}

@article{Berta2024,
  author  = {Berta, V. and others},
  title   = {A Fourth-Order Accurate Finite Volume Method for Ideal Classical and Special Relativistic MHD Based on Pointwise Reconstructions},
  journal = {Journal of Computational Physics},
  year    = {2024},
  volume  = {499},
  pages   = {112701},
  doi     = {10.1016/j.jcp.2023.112701}
}

@article{Markidis2010_iPIC3D,
  author  = {Markidis, S  and others},
  title   = {Multi-Scale Simulations of Plasma with iPIC3D},
  journal = {Mathematics and Computers in Simulation},
  volume  = {80},
  number  = {7},
  pages   = {1509--1519},
  year    = {2010},
  doi     = {10.1016/j.matcom.2009.08.038}
}

@article{Mignone2010_GLM_MHD,
  author  = {Mignone, A. and others},
  title   = {High-Order Conservative Finite Difference GLM-MHD Schemes for Cell-Centered MHD},
  journal = {Journal of Computational Physics},
  volume  = {229},
  pages   = {5896--5920},
  year    = {2010},
  doi     = {10.1016/j.jcp.2010.04.013}
}

@article{Rossazza2026_gPLUTO,
  author  = {Rossazza, M. and others},
  title   = {The PLUTO Code on GPUs: A First Look at Eulerian MHD Methods},
  journal = {Astronomy and Computing},
  volume = {55},
  pages = {101076},
  year = {2026},
  doi  = {10.1016/j.parco.2016.01.001}
}

@article{Suriano2026_LP,
title = {The PLUTO code on GPUs: Offloading Lagrangian Particle methods},
journal = {Astronomy and Computing},
volume = {55},
pages = {101088},
year = {2026},
issn = {2213-1337},
doi = {https://doi.org/10.1016/j.ascom.2026.101088},
author = {Suriano, A. and others}
}

@article{Mignone2007_PLUTO,
  author  = {Mignone, A. and others},
  title   = {PLUTO: A Numerical Code for Computational Astrophysics},
  journal = {The Astrophysical Journal Supplement Series},
  volume  = {170},
  number  = {1},
  pages   = {228--242},
  year    = {2007},
  doi     = {10.1086/513421}
}

@inproceedings{Williams2023_iPIC3D_SC23,
  author    = {Williams, J. J. and others},
  title     = {Characterizing the Performance of the Implicit Massively Parallel Particle-in-Cell iPIC3D Code},
  booktitle = {Proceedings of the International Conference for High Performance Computing,
               Networking, Storage and Analysis (SC'23)},
  year      = {2023},
  note      = {arXiv:2408.01983},
  doi       = {10.48550/arXiv.2408.01983}
}

@inproceedings{Almerol2025_Wormhole,
  author    = {Almerol, J. L and others},
  title     = {Accelerating Gravitational {N}-Body Simulations Using the {RISC-V}-Based {Tenstorrent Wormhole}},
  booktitle = {Proceedings of the SC '25 Workshops of the International Conference for High Performance Computing, Networking, Storage and Analysis},
  year      = {2025},
  pages     = {1729--1735},
  doi       = {10.1145/3731599.3767528},
  publisher = {Association for Computing Machinery},
  address   = {New York, NY, USA},
  isbn      = {9798400718717}
}

@inproceedings{simsek2024increasing,
  title={Increasing energy efficiency of astrophysics simulations through GPU frequency scaling},
  author={Simsek, O S and others},
  booktitle={SC24-W: Workshops of the International Conference for High Performance Computing, Networking, Storage and Analysis},
  pages={1826--1834},
  year={2024},
  organization={IEEE}
}

@article{silvano2025survey,
  title={A survey on deep learning hardware accelerators for heterogeneous hpc platforms},
  author={Silvano, C and others},
  journal={ACM Computing Surveys},
  volume={57},
  number={11},
  pages={1--39},
  year={2025},
  publisher={ACM New York, NY}
}

@article{elsharkawy2025integration,
  title={Integration of quantum accelerators with high performance computing—a review of quantum programming tools},
  author={Elsharkawy, A and others},
  journal={ACM Transactions on Quantum Computing},
  volume={6},
  number={3},
  pages={1--46},
  year={2025},
  publisher={ACM New York, NY}
}

@article{Hennessy2019,
author = {Hennessy, J. L. and Patterson, D. A.},
title = {A new golden age for computer architecture},
year = {2019},
issue_date = {February 2019},
publisher = {Association for Computing Machinery},
address = {New York, NY, USA},
volume = {62},
number = {2},
issn = {0001-0782},
url = {https://doi.org/10.1145/3282307},
doi = {10.1145/3282307},
journal = {Commun. ACM},
month = jan,
pages = {48–60},
numpages = {13}
}

@inproceedings{bartolini2022monte,
  title={Monte Cimone: paving the road for the first generation of RISC-V high-performance computers},
  author={Bartolini, A and others},
  booktitle={2022 IEEE 35th International System-on-Chip Conference (SOCC)},
  pages={1--6},
  year={2022},
  organization={IEEE}
}

@inbook{MarkidisGPU,
   title={Multi-GPU Acceleration of the iPIC3D Implicit Particle-in-Cell Code},
   booktitle={Computational Science – ICCS 2019},
   publisher={Springer International Publishing},
   author={Sishtla, CP and others},
   year={2019},
   pages={612–618}
}

@misc{BoellaGPU,
      title={Accelerating the Particle-In-Cell code ECsim with OpenACC}, 
      author={Boella, E and others},
      year={2026},
      eprint={2603.16624},
      archivePrefix={arXiv},
      primaryClass={physics.plasm-ph},
      url={https://arxiv.org/abs/2603.16624}, 
}

@INPROCEEDINGS{Viviani2025,
  author={Viviani, P and others},
  booktitle={2025 55th Annual IEEE/IFIP International Conference on Dependable Systems and Networks Workshops (DSN-W)}, 
  title={Assessing the Elephant in the Room in Scheduling for Current Hybrid HPC-QC Clusters}, 
  year={2025},
  volume={},
  number={},
  pages={184-187},
  doi={10.1109/DSN-W65791.2025.00059}}

@INPROCEEDINGS{Rocco_quantum,
  author={Rocco, R and others},
  booktitle={2025 IEEE International Conference on Quantum Computing and Engineering (QCE)}, 
  title={Dynamic Solutions for Hybrid Quantum-HPC Resource Allocation}, 
  year={2025},
  volume={02},
  number={},
  pages={34-40},
  doi={10.1109/QCE65121.2025.10289}}

@article{Rocco2025,
title = {To repair or not to repair: Assessing fault resilience in MPI stencil applications},
journal = {Journal of Parallel and Distributed Computing},
volume = {205},
pages = {105156},
year = {2025},
author = {Rocco, R and others}
}

@misc{brown2025riscvreadyhighperformance,
      title={Is RISC-V ready for High Performance Computing? An evaluation of the Sophon SG2044}, 
      author={Nick Brown},
      year={2025},
      eprint={2508.13840},
      archivePrefix={arXiv},
      primaryClass={cs.DC},
      url={https://arxiv.org/abs/2508.13840}, 
}

@inproceedings{ristov2016superlinear,
  title={Superlinear speedup in HPC systems: Why and when?},
  author={Ristov, S and others},
  booktitle={2016 federated conference on computer science and information systems (fedcsis)},
  pages={889--898},
  year={2016},
  organization={IEEE}
}

@InProceedings{montecimone2,
author={Venieri, E and others},
editor="Neuwirth, Sarah
and Paul, Arnab Kumar
and Weinzierl, Tobias
and Carson, Erin Claire",
title="Monte Cimone v2: HPC RISC-V Cluster Evaluation and Optimization",
booktitle="High Performance Computing",
year="2026",
publisher="Springer Nature Switzerland",
address="Cham",
pages="576--585",
isbn="978-3-032-07612-0"
}

@article{GEM,
author = {Birn, J. and others},
title = {Geospace Environmental Modeling (GEM) Magnetic Reconnection Challenge},
journal = {Journal of Geophysical Research: Space Physics},
volume = {106},
number = {A3},
pages = {3715-3719},
doi = {https://doi.org/10.1029/1999JA900449},
url = {https://agupubs.onlinelibrary.wiley.com/doi/abs/10.1029/1999JA900449},
eprint = {https://agupubs.onlinelibrary.wiley.com/doi/pdf/10.1029/1999JA900449},
year = {2001}
}

@article{almerol2026riscv_nbody,
title = {Assessing performance and porting strategies for gravitational N-body simulations on the RISC-V-based tenstorrent Wormhole™},
journal = {Astronomy and Computing},
pages = {101121},
year = {2026},
issn = {2213-1337},
doi = {https://doi.org/10.1016/j.ascom.2026.101121},
author = {Jenny Lynn Almerol and others}
}
\end{document}